# A Comparative Analysis of Data Mining Tools in Agent Based Systems

Sharon Christa[1]   K. Lakshmi Madhuri[2]   V. Suma[3]
[1.] Post Graduate Programme, Dept of ISE, Dayananda Sagar College of Engineering, Bangalore, India.
[2, 3.] Research and Industry Incubation Centre, Dayananda Sagar Institutions, Bangalore, India.
[1.] sharonchrista@gmail.com    [2.] madhuri.vethamoorthy@gmail.com    [3.] sumavdsce@gmail.com

**Abstract**
World wide technological advancement has brought in a widespread change in adoption and utilization of open source tools. Since, most of the organizations across the globe deal with a large amount of data to be updated online and transactions are made every second, managing, mining and processing this dynamic data is very complex. Successful implementation of the data mining technique requires a careful assessment of the various tools and algorithms available to mining experts. This paper provides a comparative study of open source data mining tools available to the professionals. Parameters influencing the choice of apt tools in addition to the real time challenges are discussed. However, it is well proven that agents aid in improving the performance of data mining tools. This paper provides information on an agent-based framework for data preprocessing with implementation details for the development of better tool in the market. An integration of open source data mining tools with agent simulation enable one to implement an effective data pre processing architecture thereby providing robust capabilities of the application which can be upgraded using a minimum of pre planning requirement from the application developer.

**Introduction**

The recent advance in technology allows organizations to store complex data in various locations and also allows data updation in every second. The main challenge of any organization is to effectively manage large updated data online. One of the influencing features that have constantly hindered the Data Stream Management System is its inability to simultaneously query both live and archival data [1]. Data Mining is the process that helps to make use of the data in various databases and find new patterns in it. Data mining model is successfully able to solve emergent problems such as the discovery of patterns and knowledge in uncertain, high-frequency, organizational, or behavioral data, including data generated and stored in various locations systems [3]. Applying data mining to acquire updated knowledge leads to complexity in data processing, managing and mining [2]. Commonly used open source data mining tools are Weka, RapidMiner, TANAGRA, Orange, and KNIME. Data mining tools provide us with user friendly interface for data analysis and its main features are Handling Complicated Problem, Discovering Unknown Patterns, Skill Required in Working with the Tool, Scalability, Data mining tools should be capable of handling large amount of datasets, Cost [11]. According to Computer system environment for data mining tools require a client server environment [9]. The data mining tools has to be analyzed based on parameters like Product track record, Vendor Viability, Breadth of Data Mining algorithms, Compatibility with a specific computer environment, Ease of use and ability to handle large databases

Data mining experts of Pharmine Company has summarized a report on comparison of data mining tools, which evaluates various data mining tools like KNIME, Rapid Miner, Weka, Tanagra and Orange [10]. Table 1 depicts the result chart of the Data Mining Tool Comparison developed by Pharmine research is given below.

| Procedure | KNIME | RapidMiner | Weka | TANAGRA | Orange |
|---|---|---|---|---|---|
| Partitioning of dataset to training and testing sets | Have with limited partitioning abilities | Have with limited partitioning abilities | Have with limited partitioning abilities | Have with limited partitioning abilities | Have with limited partitioning abilities |
| Descriptor Scaling | Have the facility | Have the facility | Does not have the facility to save parameters for scaling to apply to future datasets | Does not have the facility to save parameters for scaling to apply to future datasets | Does not have the facility of scaling |
| Descriptor Selection | Has no wrapper methods | Have the facility | Have the facility but not the part of knowledge flow | Have wrapper methods that is valid only for logistic regression | Has no wrapper methods |
| Parameter optimization of machine learning/statistical methods | Does not have automatic facility | Has the facility | Does not have automatic facility | Does not have automatic facility | Does not have automatic facility |
| Model validation using cross-validation and/or independent validation set | Have only limited error measurement methods | Has the full facility | Have the facility but is not capable of saving the model so have to rebuild model for every future data set | It does not have the capability of validating independent validation set | Have the facility but is not capable of saving the model so have to rebuild model for every future data set |

Table 1: Comparative analysis of data mining tools

Noisy data is one of the major challenges in data mining processes. The current data mining tools do not support real time modeling where new model or knowledge can be acquired when new data is added. In practice, a data miner will carry out remodeling to ensure new knowledge is obtained. This limitation leads to hitch in many applications such as in medical, stock exchange and finance. These domains are critical in having updated knowledge since it affects the human life and business performances [4]. Successful implementation of the data mining effort requires a careful assessment of the various tools and algorithms available. However, current era of information age has forced the industries to deal with dealing with gigabytes and terabytes of database in lieu of megabytes of database [7]. As data sets increase in size, data mining tools become less and less efficient. This is an important feature for a data mining tool which motivates us to analyze the performance of the data mining tools. In order to incorporate the most recent knowledge model in the current data mining tools is expensive. With arrival of every data, it is required to clean and mine dynamically. Additionally, the users have to decide an appropriate time to perform data mining which is again time intensive factor [4]. The available data mining tools require expert users to carry out experiments in order to get a useful knowledge model and it should be performed in a way where data is treated equally. The choice of techniques depends on the knowledge of the user. Existing data mining tools do not support arrival of new data and re-model the knowledge while some significant valuable data is added. The new data added should be mined again to get the updated knowledge model. Hence, currently available data mining tools are not suitable for novice users [4].

Agent based systems are the outstanding approach to overcome the drawbacks cited above. In recent years, agents have become popular paradigm in computing because of its autonomous, flexible, adaptive and intelligent characteristics [5]. Intelligent agents behave rationally [6]. The intelligent agents do the work with human intelligence but may not behave like human beings. The agent which is involved in processing of mining performs productive task, retrieves useful knowledge with less noise and reduced processing time when compared to the normal mining tools [4]. Potential features of an agent based data mining tools include:
 a) Propose the processing techniques most suitable to the data
 b) Preprocess incoming new data according to user profile
 c) Share mining experience
 d) Suggesting possible knowledge that can be extracted from the data with the help of the experience shared
 e) Suitable for novice user

**Existing Agent Simulation Tools**

In this section we would like to discuss the popular agent simulation tools like Aglets, JATLite, FTP Software Agent and Voyager.

Aglets are Java-based autonomous agents developed by IBM, which provide the basic capabilities required for mobility and has a globally unique name. A travel itinerary is used to specify the destinations to which the agent must travel and what actions it must take at each location. In order for an aglet to run on a particular system, the target system must be running an aglet host application which provides a platform-neutral execution environment for the aglet. The aglet workbench includes a configurable Java security manager. Aglets can communicate using a whiteboard that allows agents to collaborate and share information asynchronously. Synchronous and asynchronous message passing is also supported for aglet communication. Aglets are streamed using standard Java serialization or externalization. A network agent class loader is supplied which allows an aglet's byte code stream and state to travel across a network [13].

Java Agent Template Lite (JATLite) is a set of light-weight Java packages being developed at Stanford University that can be used to build multiagent systems. It is a layered architecture which provides a different communication protocol at each layer. The JATLite framework is intended for developing typed-message, autonomous agents that communicate using a peer-to-peer protocol. Both synchronous and asynchronous message passing are supported. Messages can be delivered through polling or message queuing. The framework provides additional security which checks the agent name and password for a more secure connection [14].

FTP Software Agent Technology is Java-based software designed to manage heterogeneous networks across the Internet using agent technology. The agents are autonomous and mobile, and can move to any system in the network which has an Agent Responder installed. As the agent moves from system to system, its tasks may change, depending on the environment of the system it is visiting.
The agents can interact with other agents or with the user, as needed [15]. But FTP agents do not require any user interaction—based on push technology, they can move from system to system, respond to events, and perform tasks according to criteria predefined by the user. An Agent Manager is responsible for launching the agent.

Voyager, from ObjectSpace, Inc., is an agent-enhanced Object Request Broker (ORB) coded in Java. An ORB provides the capability to create objects on a remote system and invoke methods on those objects [16]. Voyager augments the traditional ORB with agent capabilities. Voyager agents have mobility and autonomy which is provided in the base class, Agent. An Agent can move itself from one location to another and can leave behind a forwarding address with a secretary so that future messages can be forwarded to its new location. Specialized agents, called Messengers, are used to deliver messages.

Messages can be synchronous, one-way i.e. similar to asynchronous, or future, which are asynchronous but return a placeholder that can be used to retrieve a return value at a later time. An agent's itinerary instructs the agent as to what operations it needs to perform at each location. Voyager has a security manager which can be used to restrict the operations an agent can perform.

**Architecture Development**

With the emergence of online transactions, World Wide Web, data streams etc., large amount of data is created and are being stored in databases, which is one of the reason for complexity of data management. Novice users may find it difficult to determine the best technique for pre-processing their data to perform data mining. Hence, experts are required to identify the best-suited pre-processing technique [4]. At present, neither data mining tools are efficient in handling dynamic and complex data nor do the users have sufficient knowledge in pre processing the data in terms of data mining domain. However, with the integration of agent based data mining system, the agent determines the technique and the parameters that provide the best model for good decision making. Since, the current mining tools are domain specific, this research focused us to propose a generic architecture that can pre-process data using agents of any domain of application. In addition, even a novice user can use the proposed architecture [12].

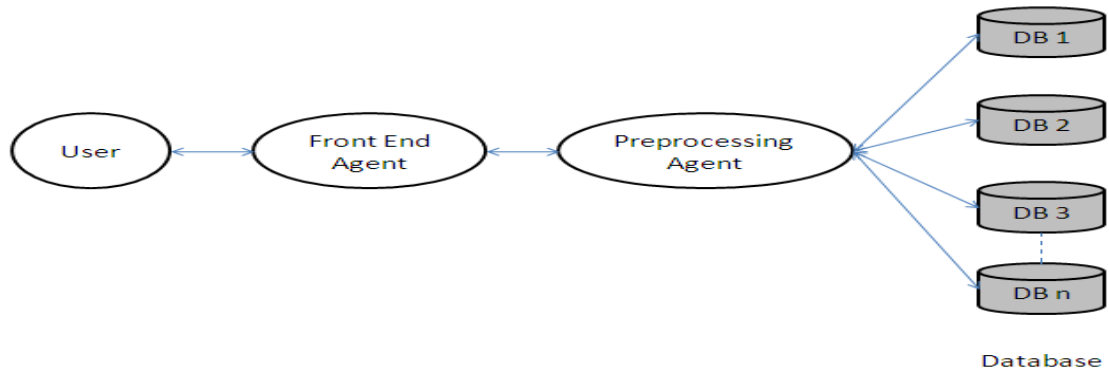

Figure 1: Proposed Pre-processing Architecture.

The proposed architecture in figure 1 is the basic model of Pre-processing architecture. The framework is designed with five major agents which includes User Interface Agent (UI Agent), Coordinator Agent, Clean Agent, Transformation Agent and Reduction Agent. Figure 2 depicts the aforementioned design of pre-processing architecture. The responsibilities and capabilities are specific of each agent. The User Interface Agent provides the user interface with the system. It provides solution analysis autonomously and helps the user to give queries according to the requirement of user. Coordinator agent is responsible for the coordination of all the tasks that is performed in the system. It determines the pre-processing method to be used based on the data mining task given by the user which is generated according to the Meta knowledge which the agent maintains. It has access to the data repository that can be updated dynamically and provides data to the other agents. Coordinator agent also provides adaptive profiling user data and checks to identify the data types and attributes in the database. It identifies the problems the data has and save the knowledge and corresponding preprocessing technique that is best for it. Clean Agent handles the missing and noisy data using the techniques in the data dictionary. Transformation Agent is used to transform the data into appropriate forms for mining. The role of reduction agent is to reduce the size of the data by using selected discretization techniques.

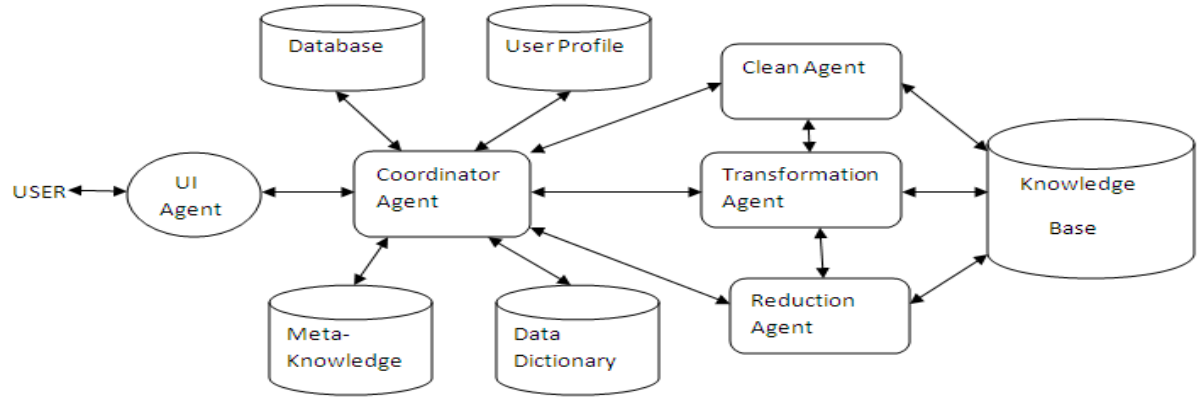

Figure 2: Pre-processing Architecture

**Implementation of the Agent Based Preprocessing Architecture**

The proposed architecture model is a client server model with a basic Model View Controller Architecture. The above proposed architecture can be implemented in different ways like implementing both agent and application or to modify the existing code of an application to enable the necessary communication. The preprocessing methods are all available in various open source data mining tools and open source agent simulation tools.

The agent framework which we proposed will have the following requirements
a) Ability to add intelligence to applications.
b) Intelligent agent framework is practical to solve real-world problems.
c) Architecture must be flexible enough to support any applications.
d) Intelligent agents increase the functionality of the application and can communicate with each other and other applications.
e) The agents can call the shots and monitor and drive the applications.

**Functional Specification of Agent Framework**

The functionality of Agent Framework always contain the following specifications
i. It must be easy to add an intelligent agent to an existing application.
ii. A graphical construction tool must be available to compose agents out of other Java components and other agents.
iii. The agents must support a relatively sophisticated event processing capability. Agent will need to handle events from the outside world, other agents, and signal events to outside applications.
iv. Domain knowledge can be added to agent using if-then rules, and support forward and backward rule-based processing with sensors and effectors.
v. The agents must be able to learn to do classification, clustering, and prediction using learning algorithms.
vi. Multi agent applications must be supported using a KQML-like message protocol.
vii. The agent should be persistent. That is, once an agent is constructed, there must be a way to save it in a file and reload its state at a later time.

Due to the availability of open source data mining tools and agent simulation tools for the implementation of the architecture, it is now possible to append an agent to the existing application, thereby extending the basic capabilities of the application which requires a minimum of pre planning of the application developer.

**Conclusion**

State of the art advancement in technology enables organizations to store, process and update huge and complex data dynamically. The development and application of data mining techniques requires the use of right choice of software tools. Further, current data mining tools are expensive to get the updated knowledge model. This paper provides information on two aspects of data mining tools namely to elucidate the comparative analysis of various open source data mining tools and to put forth the challenges which exist in the existing data mining tools. However, agents are known to aid in improving the performance of data mining tools. This paper has therefore proposed to integrate existing data-mining tool with agents in order to implement an effective data preprocessing architecture. The functional specifications elucidated here enable the application developer to accurately analyze, evaluate and develop the data pre processing tool for improved data management.